\documentclass[12pt]{article}
\usepackage[dvips]{graphics}

\def\beq   {\begin{equation}}
\def\eeq   {\end{equation}}
\def\beqd  {\begin{displaymath}}
\def\eeqd  {\end{displaymath}}
\def\beqaa {\begin{eqnarray}}
\def\eeqaa {\end{eqnarray}}

\def\noi {\noindent}

\def\ti  {\tilde}

\def\sn  {\ti \nu}
\def\sl  {\ti \ell}

\def\nt  {\tilde\chi^0}
\def\ch  {\tilde\chi^\pm}
\def\chp {\tilde\chi^+}

\def\a   {\alpha}
\def\b   {\beta}
\def\t   {\theta}

\def\sz{\ifmmode{\tilde{\chi}^0} \else{$\tilde{\chi}^0$} \fi}
\def\sw{\ifmmode{\tilde{\chi}} \else{$\tilde{\chi}$} \fi}

\newcommand{\gsim}{\;\raisebox{-0.9ex}
           {$\textstyle\stackrel{\textstyle >}{\sim}$}\;}
\newcommand{\lsim}{\;\raisebox{-0.9ex}{$\textstyle\stackrel{\textstyle<}
           {\sim}$}\;}

\begin{document}
%------------------------------------------------------------------------
\pagestyle{empty}

\vspace*{1.4cm}

\begin{center}

{\Large {\bf
Impact of slepton generation mixing on the search for sneutrinos}
}\\

\vspace{10mm}

{\large 

A.~Bartl$^a$, K.~Hidaka$^b$, K.~Hohenwarter-Sodek$^a$, T.~Kernreiter$^a$,
W.~Majerotto$^c$ and W.~Porod$^d$

}

\vspace{6mm}

\begin{tabular}{l}
$^a${\it  Faculty of Physics, University of Vienna, 
Boltzmanngasse 5, A-1090 Wien, Austria}\\
$^b${\it Department of Physics, Tokyo Gakugei University, Koganei,
Tokyo 184--8501, Japan}\\
$^c${\it Institut f\"ur Hochenergiephysik der \"Osterreichischen Akademie 
der Wissenschaften,} \\
{\it A-1050 Vienna, Austria}\\
$^d${\it Institut f\"ur Theoretische Physik and Astrophysik, Universit\"at 
W\"urzburg,} \\
{\it D-97074 W\"urzburg, Germany}
\end{tabular}

\end{center}

%\vspace{3cm}
\vfill

\begin{abstract} 
%
%We perform a systematic study of sneutrino decays in the Minimal Supersymmetric 
%Standard Model (MSSM) with slepton generation mixing. 
We perform a systematic study of sneutrino production and decays in the Minimal 
Supersymmetric Standard Model (MSSM) with slepton generation mixing. 
We study both fermionic decays like 
$\tilde\nu\to\ell^- \tilde\chi^+, \nu\tilde\chi^0$ and bosonic decays such as 
$\tilde\nu\to  \tilde\ell^- H^+, \tilde\ell^- W^+$.
%We show that the effect of slepton generation mixing 
%on the sneutrino decay branching ratios
%can be quite strong in a significant part of the 
We show that the effect of slepton generation mixing 
on the sneutrino production cross sections and its decay branching ratios 
can be quite large in a significant part of the 
MSSM parameter space despite the very strong experimental limits 
on lepton flavour violating processes. This could have an important impact 
on the search for sneutrinos and the determination of the MSSM parameters 
at future colliders, such as LHC, ILC, CLIC and muon collider.
\end{abstract}

\newpage
\pagestyle{plain}
\setcounter{page}{2}

% --- introduction ---
%-----------------------------------------------------------------------------
\section{Introduction}
%-----------------------------------------------------------------------------

In the Minimal Supersymmetric Standard Model (MSSM) \cite{Kane}, 
supersymmetric (SUSY) partners of all Standard Model 
(SM) particles with masses less than ${\mathcal O}$(1 TeV) are introduced. 
In this way, SUSY solves the problems of hierarchy, fine-tuning 
and naturalness of the SM. Hence, the discovery of the SUSY partners 
and the study of their properties play an important role at future 
%colliders, such as the Large Hadron Collider (LHC) \cite{LHCTDR}, 
%$e^+e^-$ linear colliders (ILC and CLIC) \cite{LC,Accomando:2004sz}, and 
%$\mu^+\mu^-$ collider \cite{MuonCollider}. They will extend the discovery 
colliders, such as the Large Hadron Collider (LHC), 
$e^+e^-$ linear colliders (ILC and CLIC), and 
$\mu^+\mu^-$ collider. They will extend the discovery 
potential for SUSY particles to the TeV mass range, and allow for a precise 
determination of the SUSY parameters.

In this letter we focus on the sneutrinos, the SUSY partners of neutrinos. 
Systematic studies of sneutrino decays in the MSSM including the 
effects of CP violation have been performed already \cite{Bartl:sleptondecay,Bartl:2002bh}.
In these studies it has been assumed that individual lepton flavour is 
conserved in the slepton sector, which is suggested by the so far
negative searches for lepton flavour 
violating (LFV) processes, e.g. $\mu\to e~ \gamma$.
However, such an assumption may be too restrictive.
Furthermore, it has been shown that LFV terms can be 
induced in the slepton sector even in the constrained MSSM 
(for vanishing LFV soft SUSY breaking terms at the GUT scale) 
when supplemented with right-handed neutrino superfields 
for accommodating neutrino data \cite{Borzumati:1986qx}. 
These LFV terms can give rise to LFV SUSY particle reactions at an 
observable level.
For the ILC, LFV reactions in SUSY particle productions and decays 
have been studied for the cases of two generation mixings 
$\tilde e-\tilde\mu$ \cite{Krasnikov:1995qq,Arkani-Hamed:1996au,
Feng:1998ud,Oshimo:2004wv,Deppisch:2004fz}, 
$\tilde \mu-\tilde\tau$ \cite{Guchait:2001us}, 
and $\tilde e-\tilde\tau$ \cite{Nomura:2000zb,HohenwarterSodek:2007az}. 
Moreover, the case of three generation mixing $\tilde e-\tilde\mu-\tilde\tau$
has been studied in 
\cite{Nomura:2000zb,Hirouchi:1997cy,Hisano:1998wn,Porod:2002zy,
Deppisch:2003wt,Bartl: LFV@LHC,Dedes:2007ef} 
(in \cite{Hisano:1998wn,Cheng:1997nd} also for a muon collider).
Some of the studies above, i.e. 
\cite{Nomura:2000zb,Hirouchi:1997cy,Hisano:1998wn,Deppisch:2003wt,Dedes:2007ef}, are rather 
model dependent.
In contrast, our framework is the general MSSM, where we impose no further
assumptions other than important experimental and theoretical requirements
which have to be respected by the MSSM parameters.
Furthermore, so far no systematic study of LFV in sneutrino
decays including bosonic decays has been performed.
%The aim of this article is to perform a systematic study of sneutrino decays 
The aim of this article is to perform a systematic study of sneutrino 
production and decays 
including the bosonic decay modes in the general MSSM with LFV 
in slepton sector. We study the cases of two generation mixings 
in the slepton sector, i.e. $\tilde\mu-\tilde\tau$ and 
$\tilde e-\tilde\tau$ mixings. We do not expect significant 
LFV effects in sneutrino decays in the $\tilde e-\tilde\mu$
mixing case \cite{Oshimo:2004wv}. 
%We show that the LFV decay branching ratios for both
%fermionic and bosonic sneutrino decays can be quite large 
We show that the LFV sneutrino production cross sections and 
the LFV decay branching ratios for both
fermionic and bosonic sneutrino decays can be quite large 
due to slepton generation mixing in a significant part of the MSSM parameter 
space despite the very strong experimental limits on LFV processes. 
This could have an important impact on the search for sneutrinos and the 
MSSM parameter determination at future colliders, such as LHC, ILC, 
CLIC and muon collider.

In section \ref{model} we give an account on slepton generation mixing in the 
general MSSM. In Section \ref{constraints} we list the constraints which we
impose on the model parameters. 
%Our numerical analysis on LFV 
%sneutrino decays is given in section \ref{numerics}.
Our numerical analysis on LFV 
sneutrino production and decays is given in section \ref{numerics}.
We summarize in section \ref{summary}.

%-----------------------------------------------------------------------------
\section{The model\label{model}}
%-----------------------------------------------------------------------------
  First we summarize the MSSM parameters in our analysis. 
The most general charged slepton mass matrix including left-right mixing
as well as flavour mixing in the basis of
$\sl_{0\a}=(\tilde e_L,\tilde\mu_L,\tilde\tau_L,\tilde
e_R,\tilde\mu_R,\tilde\tau_R)$, $\a=1,...,6$,  
is given by \cite{Bartl: LFV@LHC,Chung}:
\begin{equation}
M^2_{\tilde \ell} = \left(
\begin{array}{cc}
M^2_{LL} &  M^{2\dagger}_{RL} \\
M^2_{RL} &  M^2_{RR} \\
\end{array} \right)~,
\label{eq:sleptonmass}
\end{equation}
where the entries are $3 \times 3$ matrices. They are given by
\begin{eqnarray}
\label{eq:sleptonmassLL}
M^2_{LL,\alpha\beta} &=& 
M^2_{L,\alpha\beta} + m^2_Z\cos(2\b)(-\frac{1}{2}+\sin^2{\t_W})\delta_{\a\b}
+m^2_{\ell_\a}\delta_{\a\b} \, ,\\
\label{eq:sleptonmassRR}
M^2_{RR,\a\b} &=& M^2_{E,\a\b}-m^2_Z\cos(2\b)\sin^2{\t_W}\delta_{\a\b} 
+m^2_{\ell_\a}\delta_{\a\b} \, ,\\
\label{eq:sleptonmassRL}
M^2_{RL,\a\b} &=& v_1 A_{\b\a}-m_{\ell_\a}\mu^*\tan\b\delta_{\a\b} \, .
\end{eqnarray}
The indices $\a,\b=1,2,3$ characterize the flavours 
$e,\mu,\tau$, respectively.
$M^2_{L}$ and $M^2_{E}$ are the hermitean soft SUSY breaking mass matrices for
left and right sleptons, respectively. 
$A_{\a\b}$ are the trilinear soft
SUSY breaking couplings of the sleptons and the Higgs boson:
${\mathcal L}_{\rm int}=-A_{\a\b} \sl_{\b R}^\dagger \sl_{\a L} H^0_1 
 + A_{\a\b} \sl_{\b R}^\dagger \sn_{\a L} H^-_1 + \cdots$.
$\mu$ is the higgsino mass parameter.
$v_1$ and $v_2$ are the vacuum expectation values of the Higgs fields
with $v_1=\langle H^0_1\rangle$, $v_2=\langle H^0_2\rangle$, 
and $\tan\b\equiv v_2/v_1$. 
We work in a basis where the Yukawa coupling
matrix $Y_{E,\a\b}$ of the charged leptons is real and flavour
diagonal with $Y_{E,\a\a}=m_{\ell_\a}/v_1=\frac{g}{\sqrt{2}}
\frac{m_{\ell_\a}}{m_W\cos\b}~
(\ell_\a=e,\mu,\tau)$, with $m_{\ell_\a}$ being the physical lepton masses and 
$g$ the SU(2) gauge coupling.
The physical mass eigenstates $\sl_i$, $i=1,...,6$, are given by 
$\sl_i = R^{\sl}_{i\a} \sl_{0\a}$. 
The mixing matrix $R^{\sl}$ and the physical mass eigenvalues
are obtained by an unitary transformation 
$R^{\sl}M^2_{\sl}R^{\sl\dagger}=
{\rm diag}(m^2_{\sl_1},\dots,m^2_{\sl_6})$, 
where $m_{\sl_i} < m_{\sl_j}$ for $i<j$.
Similarly, the mass matrix for the sneutrinos, in the basis
$\sn_{0\a}=(\tilde\nu_{eL},\tilde\nu_{\mu L},\tilde\nu_{\tau L})\equiv
(\tilde\nu_{e},\tilde\nu_{\mu},\tilde\nu_{\tau})$, reads
\begin{eqnarray}
M^2_{\sn,\a\b} &=&  M^2_{L,\a\b} 
+ \frac{1}{2} m^2_Z\cos(2\b)\delta_{\a\b} 
\qquad (\alpha,\beta=1,2,3)~,
\label{eq:sneutrinomass}
\end{eqnarray}
where the physical mass eigenstates are given by
$\sn_i = R^{\sn}_{i\a}\sn_{0\a}$, $i=1,2,3$,
($m_{\sn_1} < m_{\sn_2} < m_{\sn_3}$).

The properties of the charginos $\ch_i$ ($i=1,2$, $m_{\ch_1}<m_{\ch_2}$) 
and neutralinos $\nt_k$ ($k=1,...,4$, $m_{\nt_1}< ...< m_{\nt_4}$)  
are determined by the parameters $M_2$, $M_1$, $\mu$ and $\tan\b$, 
where $M_2$ and $M_1$ are the SU(2) and U(1) gaugino masses, respectively. 
Assuming gaugino mass unification we take $M_1=(5/3)\tan^2\t_W M_2$. 

\noi
The possible fermionic and bosonic two-body decay modes of sneutrinos are 
\beqaa
  \sn_i  &\longrightarrow & \nu \nt_j,~ \ell^-_\a \chp_k~, \\
   \sn_i  &\longrightarrow & \sl^-_j W^+,~ \sl^-_j H^+~.     \label{eq:Bmode}
\eeqaa
Note that the neutrino flavour cannot be 
discriminated in high energy collider experiments.
The bosonic decays in (\ref{eq:Bmode}) are 
possible if the mass splitting between sneutrinos and sleptons is 
sufficiently large. 
We have used the formulas for the partial decay widths as given
in \cite{Bartl:2002bh} by appropriately modifying the couplings to inculde
flavour violation as given, for instance, in \cite{Chung}.

%-------------------------------------------------------------------
\section{Constraints\label{constraints}}
%-------------------------------------------------------------------
In our analysis, we impose the following conditions on the MSSM parameter space 
in order to respect experimental and theoretical constraints:

\renewcommand{\labelenumi}{(\roman{enumi})} 
% set counter to small roman numbers
\begin{enumerate}
  \item The vacuum stability conditions \cite{Dimopoulos}:
        $|A_{\a\a}|^2 < 3\,Y_{E,\a\a}^2(M^2_{L,\a\a} + M^2_{E,\a\a} + m^2_1)$, and 
        $|A_{\a\b}|^2 < Y_{E,\gamma\gamma}^2(M^2_{L,\a\a} + M^2_{E,\b\b} + m^2_1)$,
        $(\a\neq\b;\gamma=Max(\a,\b);\a,\b=1,2,3=e,\mu,\tau)$ 
        where 
        $m^2_1=(m_{H^+}^2 + m_Z^2 \sin^2\t_W)\sin^2\b-\frac{1}{2}\,m_Z^2$. 
  \item Experimental limits on the LFV lepton decays:  
        $B(\mu^- \to e^- \gamma) < 1.2 \times 10^{-11}$ (90\% CL) 
                                   \cite{Brooks},
        $B(\tau^- \to \mu^- \gamma) < 4.5 \times 10^{-8}$ (90\% CL) 
                                      \cite{Belle: ICHEP2006},
        $B(\tau^- \to e^- \gamma) < 1.1 \times 10^{-7}$ (90\% CL) 
                                    \cite{Babar: 2006},
        $B(\mu^- \to e^- e^+ e^-) < 1.0 \times 10^{-12}$ (90\% CL) 
                                    \cite{SINDRUM},
%        $B(\tau^-\to \mu^-\mu^+\mu^-) < 1.9 \times 10^{-7}$ (90\% CL) 
%                                        \cite{Aubert:2003pc},
%        $B(\tau^-\to e^-e^+e^-) < 2.0 \times 10^{-7}$ (90\% CL) 
%                                        \cite{Aubert:2003pc}.
        $B(\tau^-\to \mu^-\mu^+\mu^-) < 3.2 \times 10^{-8}$ (90\% CL) 
                                        \cite{BELLE: tau to 3l},
        $B(\tau^-\to e^-e^+e^-) < 3.6 \times 10^{-8}$ (90\% CL) 
                                        \cite{BELLE: tau to 3l}.
  \item Experimental limits on SUSY contributions to anomalous magnetic 
        moments of leptons \cite{PDG,Jegerlehner}
\footnote{
      For anomalous magnetic moments of electron and tau lepton 
      ($a_e$ and $a_\tau$) we take an experimental error range at 95\% CL 
      for the SUSY contributions $\Delta a_e^{SUSY}$ and 
      $\Delta a_\tau^{SUSY}$ using the data of \cite{PDG}. 
      Also for the limit on $\Delta a_\mu^{SUSY}$, we allow for 
      an error at 95\% CL for the difference 
      between the experimental measurement and the SM prediction 
      \cite{Jegerlehner}.}:
        $|\Delta a_e^{SUSY}|< 7.4 \times 10^{-12}$ (95\% CL),
        $|\Delta a_\mu^{SUSY} - 287 \times 10^{-11}|<178 \times 
         10^{-11}$ (95\% CL), $|\Delta a_\tau^{SUSY}| < 0.033$ (95\% CL).
  \item The LEP limits on SUSY particle masses 
        \cite{{LEP: ICHEP2002},{LEP: Higgs Working Group}}:
        $m_{\ch_1} > 103$ GeV, $m_{\nt_1} > 50$ GeV,
        $m_{\sl_1} > 100$ GeV, $m_{\sl_1} > m_{\nt_1}$,
        $m^2_{H^+} \gsim m^2_{A} + m^2_{W} > (93~{\rm GeV})^2 + (80~{\rm GeV})^2 =(123~{\rm GeV})^2$. 
  \item The limit on $m_{H^+}$ and $\tan\b$ from the experimental data on 
        $B(B_u^- \to \tau^- {\bar\nu}_\tau)$ \cite{BELLE: Btaunu}
\footnote{
The Babar experiment also has obtained similar data on 
$B(B_u^+ \to \tau^+ \nu_\tau)$ \cite{BABAR: Btaunu} (see also 
\cite{Hou: SUSY2007}). We have checked that our scenarios studied 
in this article are allowd also by this data.
}:
        $|R_{B\tau\nu}^{SUSY} - 1.13|<1.04$ (95\% CL)
        with $R_{B\tau\nu}^{SUSY} \equiv B^{SUSY}(B_u^- \to \tau^- {\bar\nu}_\tau) / 
                 B^{SM}(B_u^- \to \tau^- {\bar\nu}_\tau)   
               \simeq (1 - (\frac{m_{B^+}\tan\b}{m_{H^+}})^2)^2$.

\end{enumerate}
We use the one-loop formulas from \cite{Hisano:1995cp,Bartl:2003ju} for 
the computation of the SUSY contributions to the observables in conditions (ii) (for 
$\ell^- \to \ell'^- \gamma$) and (iii). 
It has been shown that in general the limit on the
$\mu^-- e^-$ conversion rate is respected if the
limit on $\mu\to e~ \gamma$ is fulfilled \cite{Hisano:1995cp}.
For the calculation of the LFV three-body decay modes in (ii)
we use the formulas given in \cite{Arganda}.
When calculating $B^{SUSY}(B_u^- \to \tau^- {\bar\nu}_\tau)$ in (v) we use the
tree-level formula \cite{Hou:1992sy}, where only the $W^-$ and $H^-$ exchanges 
%contribute, as we do not specify the squark-gluino sector.
contribute, as we do not specify the squark-gluino sector. 
%{\bf
There are no further constraints on the basic MSSM parameters in our study 
from the other observables in the B-meson sector, like $B(b \to s \gamma)$, 
$B(B^0_s \to \mu^+ \mu^-)$ and $\Delta M_{B_s}$, since the slepton sector is 
essentially independent of the squark-gluino sector in our framework of 
the general MSSM.
%}

Condition (i) is a necessary condition for the tree-level 
vacuum stability \cite{Dimopoulos}. 
This condition strongly constrains the trilinear couplings
$A_{\alpha\beta}$, especially for small $\tan\beta$ where
the lepton Yukawa couplings $Y_{E,\a\a}$ are small.  
For the limits on the lepton flavour 
mixing parameters $A_{\a\b}$ $(\a \neq \b)$ we find that now the 
experimental limits on them from LFV lepton processes in (ii) can be 
stronger than the theoretical limits from the vacuum stability condition (i) 
depending on parameter regions (see \cite{Dimopoulos}).
(ii) strongly constrains the lepton flavour mixing parameters; e.g. 
in case of $\tilde\mu-\tilde\tau$ mixing the limit on $B(\tau^- \to \mu^- \gamma)$ 
strongly constrains the $\tilde\mu-\tilde\tau$ mixing parameters $M^2_{L,23}, 
M^2_{E,23}$, $A_{23}$ and $A_{32}$. 
%(v) strongly constrains $m_{H^+}$ and $\tan\b$, where, however, 
%the restrictions are less severe
%given the recent measured value for $B(B_u^- \to \tau^- {\bar\nu}_\tau)$ %\cite{Ikado:2006un}. 
(v) strongly constrains $m_{H^+}$ and $\tan\b$.
The limit on $\Delta a_\mu^{SUSY}$ in (iii) is also important, e.g. it disfavours 
negative $\mu$ especially for large $\tan\b$.

%-------------------------------------------------------------------
\section{Numerical results\label{numerics}}
%-------------------------------------------------------------------

We take $\tan\b, m_{H^+}, M_2, \mu, M^2_{L,\a\b}, M^2_{E,\a\b}$, and $A_{\a\b}$ 
as the basic MSSM parameters at the weak scale. We assume them to be real. 
%The basic MSSM parameters at the weak scale are $\tan\b, m_{H^+}, M_2, \mu, 
%M^2_{L,\a\b}, M^2_{E,\a\b}$, and $A_{\a\b}$. We assume them to be real. 
The LFV parameters are $M^2_{L,\a\b}, M^2_{E,\a\b}$, and $A_{\a\b}$ with $\a \neq \b$.
First we study in detail $\tilde\mu-\tilde\tau$ mixing with the
parameters as given in Table \ref{tab1}. 
This scenario is within the reach of LHC and ILC and
satisfies the conditions (i)-(v).
For the branching ratios of the LFV leptonic two-body decays we obtain
$B(\mu^- \to e^-  \gamma)=1.59\times10^{-14}$, 
$B(\tau^- \to \mu^-  \gamma)=2.35\times10^{-8}$, and
$B(\tau^- \to e^-  \gamma)=2.71\times10^{-15}$.
The branching ratios for the LFV leptonic three-body decays are calculated to
be $B(\mu^-\to e^-e^+e^-)=9.7\times 10^{-17}$,
$B(\tau^-\to e^-e^+e^-)=2.9\times 10^{-17}$, and
$B(\tau^-\to \mu^-\mu^+\mu^-)=5.4\times 10^{-11}$.
The SUSY contributions to the anomalous magnetic moments of the leptons
are calculated to be $\Delta a_e^{SUSY}=2.82\times10^{-14}$, 
%are calculated to $\Delta a_e^{SUSY}=2.82\times10^{-14}$, 
$\Delta a_\mu^{SUSY}=1.24\times10^{-9}$, and 
$\Delta a_\tau^{SUSY}=3.40\times10^{-7}$.
The resulting chargino and neutralino masses are given by
$m_{\ch_1}=147$~GeV, $m_{\ch_2}=661$~GeV and 
$m_{\nt_1}=138$~GeV, $m_{\nt_2}=155$~GeV, $m_{\nt_3}=331$~GeV,
$m_{\nt_4}=661$~GeV,
respectively. 
In Table \ref{tab2} the slepton mass spectrum and the corresponding 
decompositions in flavour eigenstates are given.

\begin{table}[t]
\begin{center}
\begin{tabular}{|c|c|c|c|} \hline
 
  \multicolumn{1}{|c|}{$M_2$} 
& \multicolumn{1}{c|}{$\mu$} 
& \multicolumn{1}{c|}{$\tan\beta$} 
& \multicolumn{1}{c|}{$m_{H^+}$} \\\hline\hline
 
  \multicolumn{1}{|c|}{650} 
& \multicolumn{1}{c|}{150} 
& \multicolumn{1}{c|}{20} 
& \multicolumn{1}{c|}{150} \\\hline

\end{tabular}
\begin{tabular}{|c||c|c|c|} \hline
 $M^2_{L,\alpha\beta}$
& \multicolumn{1}{c|}{\scriptsize{${\beta=1}$}} 
& \multicolumn{1}{c|}{\scriptsize{$\beta=2$}} 
& \multicolumn{1}{c|}{\scriptsize{$\beta=3$}} \\\hline\hline
 \scriptsize{$\alpha=1$}
& \multicolumn{1}{c|}{$(430)^2$} 
& \multicolumn{1}{c|}{1} 
& \multicolumn{1}{c|}{1} \\\hline

 \scriptsize{$\alpha=2$}
& \multicolumn{1}{c|}{1} 
& \multicolumn{1}{c|}{$(410)^2$} 
& \multicolumn{1}{c|}{$(61.2)^2$} \\\hline

 \scriptsize{$\alpha=3$}
& \multicolumn{1}{c|}{1} 
& \multicolumn{1}{c|}{$(61.2)^2$} 
& \multicolumn{1}{c|}{$(400)^2$} \\\hline
\end{tabular}
\vskip0.2cm
\begin{tabular}{|c||c|c|c|} \hline
 $A_{\alpha\beta}$
& \multicolumn{1}{c|}{\scriptsize{${\beta=1}$}} 
& \multicolumn{1}{c|}{\scriptsize{$\beta=2$}} 
& \multicolumn{1}{c|}{\scriptsize{$\beta=3$}} \\\hline\hline
 \scriptsize{$\alpha=1$}
& \multicolumn{1}{c|}{0} 
& \multicolumn{1}{c|}{0} 
& \multicolumn{1}{c|}{0} \\\hline

 \scriptsize{$\alpha=2$}
& \multicolumn{1}{c|}{0} 
& \multicolumn{1}{c|}{0} 
& \multicolumn{1}{c|}{25} \\\hline

 \scriptsize{$\alpha=3$}
& \multicolumn{1}{c|}{0} 
& \multicolumn{1}{c|}{0} 
& \multicolumn{1}{c|}{150} \\\hline
\end{tabular}
\begin{tabular}{|c||c|c|c|} \hline
 $M^2_{E,\alpha\beta}$
& \multicolumn{1}{c|}{\scriptsize{${\beta=1}$}} 
& \multicolumn{1}{c|}{\scriptsize{$\beta=2$}} 
& \multicolumn{1}{c|}{\scriptsize{$\beta=3$}} \\\hline\hline
 \scriptsize{$\alpha=1$}
& \multicolumn{1}{c|}{$(230)^2$} 
& \multicolumn{1}{c|}{1} 
& \multicolumn{1}{c|}{1} \\\hline

 \scriptsize{$\alpha=2$}
& \multicolumn{1}{c|}{1} 
& \multicolumn{1}{c|}{$(210)^2$} 
& \multicolumn{1}{c|}{$(22.4)^2$} \\\hline

 \scriptsize{$\alpha=3$}
& \multicolumn{1}{c|}{1} 
& \multicolumn{1}{c|}{$(22.4)^2$} 
& \multicolumn{1}{c|}{$(200)^2$} \\\hline
\end{tabular}
%\\[0.5ex]
\vskip0.4cm
\caption{\label{tab1}
MSSM parameters in our $\tilde\mu-\tilde\tau$ mixing scenario.
All mass parameters are given in GeV.}
\end{center}
\end{table}

\begin{table}[t]
\begin{center}
%\hspace{2mm}   
\begin{tabular}{|c||c|c|c|c|}
		 \hline
			$R^{\tilde\nu}_{i\a}$
		  & $\tilde\nu_e$ & $\tilde\nu_\mu$ & $\tilde\nu_\tau$ 
                  & Mass \\
		 \hline\hline
		  $\tilde\nu_1$  & 0.0 & -0.36 & 0.93 & 393 \\
		  $\tilde\nu_2$  & 0.0 & 0.93 & 0.36 & 407 \\
		  $\tilde\nu_3$  & 1.0 & 0.0 & 0.0 & 425 \\
		 \hline
		 \end{tabular} 
\hspace{1mm}   
                 \begin{tabular}{|c||c|c|c|c|c|c|c|}
		 \hline
                 $R^{\tilde\ell}_{i\a}$ & 
                  $\tilde{e}_L$ & $\tilde{\mu}_L$ & $\tilde{\tau}_L$ &
                  $\tilde{e}_R$ & $\tilde{\mu}_R$ & $\tilde{\tau}_R$ &
                  Mass \\
		 \hline\hline
                 $\tilde\ell_1$ & 0.0 & -0.003 & 0.033 & 0.0 & -0.12 & 0.99
                 & 204 \\ 
                 $\tilde\ell_2$ & 0.0 & 0.002 & 0.004 & 0.0 & 0.99 & 0.12
                 & 215 \\ 
                 $\tilde\ell_3$ & 0.0 & 0.0 & 0.0 & 1.0 & 0.0 & 0.0 
                 & 234 \\
		 \hline
		 \end{tabular}
		
	\caption{\label{tab2} Sneutrino and charged slepton compositions, 
i.e. the mixing matrices $R^{\sn}_{i\a}$ and $R^{\sl}_{i\a}$ for $i=1,2,3$, and the associated mass 
                              spectra (in GeV) for the scenario of Table \ref{tab1}.}
\end{center}
\end{table}

\begin{table}[t]
\begin{center}

                \begin{tabular}{|l|l|l|}
		\hline\hline
			$B(\tilde\nu_1 \to \nu \tilde{\chi}^0_1)=0.066$
		 & $B(\tilde\nu_1 \to \nu \tilde{\chi}^0_2)=0.016$ 
                 & $B(\tilde\nu_1 \to \nu \tilde{\chi}^0_3)=0.045$ \\
		\hline
			$B(\tilde\nu_1 \to \mu^- \tilde{\chi}^+_1)=0.014$
		 & $B(\tilde\nu_1 \to \tau^- \tilde{\chi}^+_1)=0.36$ 
                 & \\
                \hline
			$B(\tilde\nu_1 \to  \tilde{\ell}^-_1 W^+)=0.015$
		 & $B(\tilde\nu_1 \to  \tilde{\ell}^-_1 H^+)=0.48$ 
                 &   \\
                \hline\hline
	$B(\tilde\nu_2 \to \nu \tilde{\chi}^0_1)=0.14$
	      &  $B(\tilde\nu_2 \to \nu \tilde{\chi}^0_2)=0.034$ 
                 & $B(\tilde\nu_2 \to \nu \tilde{\chi}^0_3)=0.13$ \\
		\hline
			$B(\tilde\nu_2 \to \mu^- \tilde{\chi}^+_1)=0.20$
		 & $B(\tilde\nu_2 \to \tau^- \tilde{\chi}^+_1)=0.12$ 
                 & $B(\tilde\nu_2 \to  \tilde{\ell}^-_1 H^+)=0.38$ \\
                \hline\hline
		\end{tabular}
	\caption{\label{tab3} Branching ratios of 
                              $\tilde{\nu}_1$ and $\tilde{\nu}_2$
                              decays in the scenario of Table \ref{tab1}.
Branching ratios smaller than 1\% are not shown.}
\end{center}
\end{table}

The $\sn_1$ and $\sn_2$ two-body decay branching ratios are displayed in Table \ref{tab3}.
As $\sn_2 \sim \sn_\mu$ and $\sl^-_1 \sim \ti\tau^-_R$, the decays 
$\sn_2 \to \tau^- \chp_1$ and $\sn_2 \to \sl^-_1 H^+$ 
are essentially LFV decays. Note that the branching ratios of these LFV decays 
are sizable in this scenario. The reason is as follows: 
The lighter neutralinos $\nt_{1,2}$ and the lighter chargino $\ch_1$ 
are dominantly higgsinos as $M_{1,2} \gg |\mu|$ in this scenario. 
Hence the fermionic decays into 
$\nt_{1,2}$ and $\chp_1$ are suppressed by the small lepton Yukawa 
couplings except for the decay into $\tau^- \chp_1$ which does not 
receive such a suppression because of the sizable $\tau$ Yukawa coupling 
$Y_{E,33}$ for large $\tan\b$. This leads to an enhancement of 
the bosonic decays into the Higgs boson $H^+$. 
Moreover the decay $\sn_2(\sim \sn_\mu) \to \sl^-_1(\sim \ti\tau^-_R)+ H^+$ 
is enhanced by the sizable trilinear $\sn_\mu-\ti\tau^+_R-H^-_1$ coupling 
$A_{23}$ (with $H^-_1=H^-\sin\b$). 
Because of the sizable $\tilde\nu_\mu-\tilde\nu_\tau$ mixing term $M^2_{L,23}$
the $\sn_2$ has a significant $\sn_\tau$ component, which 
results in a further enhancement of this decay due to the large trilinear 
$\sn_\tau-\ti\tau^+_R-H^-_1$ coupling $A_{33}$ ($=150$~GeV). \\
Similarly the decay $\sn_1(\sim \sn_\tau) \to \sl^-_1(\sim \ti\tau^-_R)+ H^+$ 
is enhanced by the large $A_{33}$.
The decays of $\sn_1(\sim \sn_\tau)$ and $\sn_2(\sim \sn_\mu)$ into 
$\sl^-_2(\sim \ti\mu^-_R)+ H^+$ are suppressed due to the vanishing trilinear 
couplings $A_{32}$ and $A_{22}$, respectively.
%The decays into $\sl^-_1 W^+$ are suppressed since 
%$\sl^-_1 \sim \ti\tau^-_R$.
%{\bf
The decays into $\sl^-_{1,2} W^+$ are suppressed since 
$\sl^-_1 \sim \ti\tau^-_R$ and $\sl^-_2 \sim \ti\mu^-_R$.
%}

In experimental searches for LFV in sneutrino decays it is important to have 
at least two different lepton-flavour modes with sizable branching ratios 
in decay of a sneutrino; e.g. {\it both} sizable $B(\sn_2 \to \mu^- \chp_1)$ 
{\it and} sizable $B(\sn_2 \to \tau^- \chp_1)$. 
To allow for an experimental distinction of the 
branching ratios of $\sn_1$, $\sn_2$ and $\sn_3$, sufficiently large 
mass differences among the three sneutrinos are 
required. However, large $\sn$ mass differences tend to lead to 
small $\sn$ generation mixing angles such as $\theta^{eff}_{23}$ (i.e. small 
LFV effects), where the effective $\sn_\mu - \sn_\tau$ mixing angle 
$\theta^{eff}_{23}$ is defined by 
$\tan(2\theta^{eff}_{23}) \equiv 2M^2_{L,23}/(M^2_{L,22}-M^2_{L,33})$ with 
$M^2_{L,22}-M^2_{L,33} \sim m^2_{\sn_2}-m^2_{\sn_1}$ for the scenario 
given in Table \ref{tab1}. 

%-------------------------------------------------------------------
\subsection{$\tilde\nu$ decay branching ratios \label{BRs}}
%-------------------------------------------------------------------
We study the basic MSSM parameter dependences of the LFV 
sneutrino decay branching ratios for the reference scenario of Table \ref{tab1}.
In Fig.1 we show contours of $\sn_2$ decay branching ratios 
in the $\mu-M_2$ plane. All basic parameters other than 
$\mu$ and $M_2$ are fixed as in our scenario defined in Table \ref{tab1}. 
We see that the LFV decay branching ratios 
$B(\sn_2 \to \tau^-  \chp_1)$ and 
$B(\sn_2 \to \sl^-_1  H^+)$ can be sizable in a significant part of 
the $\mu-M_2$ plane. 
The main reason for the increase of $B(\sn_2\to  \sl^-_1 H^+)$
in the region $M_2\gg \mu$ is that the partial widths for the decays into
$\mu^-\tilde\chi^+_1$ and $\nu\tilde\chi^0_{1,2}$ decrease for increasing $|M_2/\mu|$
as the lighter chargino/neutralino states become more and more
higgsino like.
The $\tau^-\tilde\chi^+_1$ decay mode has a different behaviour
due to the sizable $\tau$ Yukawa coupling for large $\tan\beta$.
%The $\tau^-\tilde\chi^+_1$ decay mode has a different behaviour
%due to the sizable Yukawa coupling for large $\tan\beta$.
We remark that the limit on $\Delta a_\mu^{SUSY}$ excludes 
the region with $B(\sn_2 \to \sl^-_1  H^+)\gsim 0.5$.
For the LFV $\tilde\nu_1$ decay we obtain $B(\sn_1 \to
\mu^-\tilde\chi^+_1)=(0.015,0.03,0.05)$
for $(\mu,M_2)$~(GeV) $= (200,600),(400,360),(600,280)$, respectively. \\
%For the LFV $\tilde\nu_1$ decay we obtain $B(\sn_1 \to
%\mu^-\tilde\chi^+_1)=(0.015,0.03,0.05)$
%for $(\mu,M_2)$~GeV $= (200,600),(400,360),(600,280)$, respectively.
%We remark that the limit on $\Delta a_\mu^{SUSY}$ excludes 
%the region with $B(\sn_2 \to \sl^-_1  H^+)\gsim 0.5$. \\
%
%In the following we use the quantity $R_{L23}\equiv
%M^2_{L,23}/((M^2_{L,11}+M^2_{L,22}+M^2_{L,33})/3)$ as a measure of LFV.
In the following we use the quantities $R_{L23}\equiv
M^2_{L,23}/((M^2_{L,11}+M^2_{L,22}+M^2_{L,33})/3)$ and $R_{A23} \equiv 
A_{23}/((|A_{11}|+|A_{22}|+|A_{33}|)/3)$ as a measure of LFV.
In Fig.2 we present the $R_{L23}$ dependence of $\sn_2$ decay branching 
ratios, where all basic parameters other than $M^2_{L,23}$ are 
fixed as in the scenario specified in Table \ref{tab1}. 
We see that the LFV decay branching ratios $B(\sn_2 \to \tau^-  \chp_1)$ 
and $B(\sn_2 \to \sl^-_1  H^+)$ can be large and very sensitive to $R_{L23}$. 
Note that $\sl^-_1 \sim \ti\tau^-_R$ and that the $\sn_\tau$ component in 
$\sn_2(\sim \sn_\mu)$ increases with the increase of the $\ti\nu_\mu - \ti\nu_\tau$ 
mixing parameter $M^2_{L,23}$, which explains the behaviour of the branching ratios.
Similarly we have found that $B(\sn_2 \to \sl^-_1  H^+)$ can be very sensitive 
to $R_{A23}$; this decay can be enhanced also by a sizable $A_{23}$ 
as explained above.
To exemplify this behaviour further, in Fig.3 we show the contours
of these decay branching ratios in the $R_{L23}-R_{A23}$ plane, where
%To exemplify this behaviour further, in Fig.3 we show the contours
%of these decay branching ratios in the $R_{L23}-R_{A23}$ plane
%(with $R_{A23} \equiv A_{23}/((|A_{11}|+|A_{22}|+|A_{33}|)/3)$), where
all basic parameters other than $M^2_{L,23}$ and $A_{23}$ are fixed
as in the scenario of Table \ref{tab1}.
As can be seen, these LFV decay branching ratios can be
large in a sizable region of the $R_{L23}-R_{A23}$ plane
and their dependences on $R_{L23}$ and $R_{A23}$ are quite remarkable and 
very different from each other.
%and their dependences on $R_{L23}$ and $R_{A23}$ are quite different.
Hence, a simultaneous measurement of these two branching 
ratios could play an important role in the determination of the LFV parameters 
$M^2_{L,23}$ and $A_{23}$. \\
In Fig.4 we show contours of the LFV decay branching ratio 
$B(\sn_1 \to \sl^-_2  H^+)$ in the $R_{E23}-R_{A32}$ plane for our scenario, 
where $R_{E23} \equiv M^2_{E,23}/((M^2_{E,11}+M^2_{E,22}+M^2_{E,33})/3)$ and 
$R_{A32} \equiv A_{32}/((|A_{11}|+|A_{22}|+|A_{33}|)/3)$. 
All basic parameters other than $M^2_{E,23}$ and $A_{32}$ are 
fixed as in the scenario described in Table \ref{tab1}. We find that 
this branching ratio can be large in a sizable region of the $R_{E23}-R_{A32}$ 
plane, and that it is sensitive to $R_{E23}$ and $R_{A32}$, which suggests  
that its measurement may be important for a 
determination of the LFV parameters $M^2_{E,23}$ and $A_{32}$. 
This behaviour of $B(\sn_1 \to \sl^-_2  H^+)$ 
can be explained as follows: 
the decay $\sn_1(\sim \sn_\tau) \to \sl^-_2(\sim \ti\mu^-_R) + H^+$ can 
be enhanced by a sizable trilinear $\sn_\tau-\ti\mu^+_R-H^-_1$ coupling 
$A_{32}$. Moreover, for a
sizable $\ti\mu_R - \ti\tau_R$ mixing term $M^2_{E,23}$, $\sl^-_2$ has 
a significant $\ti\tau^-_R$ component, which leads to an enhancement 
of this decay due to the large trilinear $\sn_\tau-\ti\tau^+_R-H^-_1$ 
coupling $A_{33}$. \\
%
%In Fig.5 we show a scatter plot of LFV decay branching ratios 
%$B(\sn_2 \to \sl^-_1  H^+)$ versus $B(\tau^- \to \mu^-  \gamma)$ for the scenario 
%in Table \ref{tab1} with the parameters $M_2,\mu,R_{L23},R_{E23},R_{A23}$ and $R_{A32}$
%are randomly generated in the ranges 
%$0<M_2<1000$~GeV, $|\mu|<1000$~GeV, $|R_{L23}|<0.1, |R_{E23}|<0.2, 
%|R_{A23}|<2.5$ and $|R_{A32}|<2.5$, respectively. 
In Fig.5 we show a scatter plot of LFV decay branching ratios 
$B(\sn_2 \to \sl^-_1  H^+)$ versus $B(\tau^- \to \mu^-  \gamma)$ for the scenario 
of Table \ref{tab1} where the parameters $M_2$, $\mu$, $R_{L23}$, $R_{E23}$, 
$R_{A23}$ and $R_{A32}$ are randomly generated in the ranges 
%{\bf
$0<M_2<1000 \, GeV$, $|\mu|<1000 \, GeV$, $|R_{L23}|<0.1, \, |R_{E23}|<0.2, \, 
|R_{A23}|<2.5$ and $|R_{A32}|<2.5$ (corresponding to $|M^2_{L,23}|<(131 \, GeV)^2, 
\, |M^2_{E,23}|<(96 \, GeV)^2, \, |A_{23}|<125 \, GeV$ and $|A_{32}|<125 \, GeV$). 
%}
%$0<M_2<1000$~GeV, $|\mu|<1000$~GeV, $|R_{L23}|<0.1, |R_{E23}|<0.2, 
%|R_{A23}|<2.5$ and $|R_{A32}|<2.5$. 
All parameters other than $ M_2, \mu, M^2_{L,23}, M^2_{E,23}, 
A_{23}$ and $A_{32}$ are fixed as given in Table \ref{tab1}. 
As can be seen in Fig.5, the LFV branching ratio 
$B(\sn_2 \to \sl^-_1  H^+)$ could go up to 30\%
even if the present bound on $B(\tau^- \to \mu^- \gamma)$ improves
by one order of magnitude.
For the other LFV decay branching ratios of $\sn_{1,2}$ versus 
$B(\tau^- \to \mu^-  \gamma)$ we have obtained scatter plots similar to that 
for $B(\sn_2 \to \sl^-_1  H^+)$ versus $B(\tau^- \to \mu^-  \gamma)$, with 
the upper limits of the $\sn_{1,2}$ decay branching ratios 
$B(\sn_1 \to \mu^-  \chp_1) \lsim 0.12$, $B(\sn_1 \to \sl^-_2  H^+) \lsim 0.40$, 
$B(\sn_1 \to \sl^-_2  W^+) \lsim 0.05$, $B(\sn_2 \to \tau^-  \chp_1) \lsim 0.35$, 
and $B(\sn_2 \to \sl^-_1  W^+) \lsim 0.22$.
Note that $B(\sn_1 \to \sl^-_2  H^+)$ can be very large 
due to sizable $M^2_{E,23}$, $A_{32}$ and large $A_{33}$, as explained above. 
$B(\sn_1 \to \mu^-  \chp_1)$ can be significant for sizable $M^2_{L23}$ 
and $M_2 \ll |\mu|$ where $\sn_1$ has a sizable $\sn_\mu$ component and 
$\chp_1$ is mainly gaugino-like. 
$B(\sn_2 \to \sl^-_1  W^+)$ can be sizable by the following reason: 
The decay $\sn_2(\sim \sn_\mu) \to \sl^-_1(\sim \ti\tau^-_R) + W^+$ 
can be enhanced by a sizable $\tilde\mu_L-\tilde\tau_R$ mixing parameter 
$A_{23}$ which leads to a sizable $\ti\mu_L$ component in $\sl_1$. 
This decay can also be enhanced by a sizable $\ti\nu_\mu - \ti\nu_\tau$ mixing 
parameter $M^2_{L,23}$ which leads to a sizable $\sn_\tau$ component in 
$\sn_2$; the $\sn_\tau$ component can couple to $W^+$ and to the sizable 
$\ti\tau_L$ component in $\sl_1$ which is due to the large 
$\ti\tau_L - \ti\tau_R$ mixing term $M^2_{RL,33}$ (for large $|\mu|\tan\b$), 
see Eq.~(\ref{eq:sleptonmassRL}).
Similarly $B(\sn_1 \to \sl^-_2  W^+)$ can be enhanced by a sizable $A_{32}$.

\begin{table}[t]
\begin{center}

                \begin{tabular}{|l|l|l|}
		\hline\hline
			$B(\tilde\nu_1 \to \nu \tilde{\chi}^0_1)=0.066$
		 & $B(\tilde\nu_1 \to \nu \tilde{\chi}^0_2)=0.016$ 
                 & $B(\tilde\nu_1 \to \nu \tilde{\chi}^0_3)=0.043$ \\
		\hline
			$B(\tilde\nu_1 \to e^- \tilde{\chi}^+_1)=0.014$
		 & $B(\tilde\nu_1 \to \tau^- \tilde{\chi}^+_1)=0.37$ 
                 & \\
                \hline
			$B(\tilde\nu_1 \to \tilde{\ell}^-_1 W^+ )=0.015$
		 & $B(\tilde\nu_1 \to  \tilde{\ell}^-_1 H^+)=0.48$ 
                 &   \\
                \hline\hline
        	$B(\tilde\nu_3 \to \nu \tilde{\chi}^0_1)=0.14$
	      &  $B(\tilde\nu_3 \to \nu \tilde{\chi}^0_2)=0.034$ 
                 & $B(\tilde\nu_3 \to \nu \tilde{\chi}^0_3)=0.12$ \\
		\hline
			$B(\tilde\nu_3 \to e^- \tilde{\chi}^+_1)=0.20$
		 & $B(\tilde\nu_3 \to \tau^- \tilde{\chi}^+_1)=0.12$ 
                 & $B(\tilde\nu_3 \to  \tilde{\ell}^-_1 H^+)=0.38$ \\
                \hline\hline
		\end{tabular}
	\caption{\label{tab4} Branching ratios of 
                              $\tilde{\nu}_1$ and $\tilde{\nu}_3$
                              decays in the $\ti e-\ti\tau$ mixing scenario.
Branching ratios smaller than 1\% are not shown.}
\end{center}
\end{table}

We have also studied the influence of LFV on the sneutrino decay branching ratios
for the case of $\tilde e-\tilde\tau$ mixing, where the MSSM parameters are
taken as in Table \ref{tab1} modified by the changes:
$M^2_{L,11}=(410~{\rm GeV})^2$, $M^2_{L,22}=(405~{\rm GeV})^2$, $M^2_{L,33}=(400~{\rm GeV})^2$,
$M^2_{L,23}\leftrightarrow M^2_{L,13}$ and $A_{23}\leftrightarrow A_{13}$.
Then $\ti\nu_1\sim \ti\nu_\tau$, $\ti\nu_2\approx \ti\nu_\mu$ and $\ti\nu_3\sim \ti\nu_e$,
where the corresponding masses are $m_{\sn_1}=393$~GeV, $m_{\sn_2}=400$~GeV 
and $m_{\sn_3}=407$~GeV, respectively. 
The resulting $\sn_1$ and $\sn_3$ decay branching ratios are given in
Table \ref{tab4}.
%As can be seen, these branching ratios are similar to the corresponding ones
%in the $\tilde\mu-\tilde\tau$ mixing scenario of Table \ref{tab1}, e.g.
%$B(\ti\nu_2\to \mu^-\ti\chi^+_1)$ in the $\ti\mu-\ti\tau$ mixing scenario 
%has the same value
%as $B(\ti\nu_3\to e^-\ti\chi^+_1)$ in the $\ti e-\ti\tau$ mixing scenario.
As can be seen, these branching ratios are similar to the corresponding ones
in the $\tilde\mu-\tilde\tau$ mixing scenario of Table \ref{tab1}, e.g.
$B(\ti\nu_3\to e^-\ti\chi^+_1)$ in the $\ti e-\ti\tau$ mixing scenario
has the same value as 
$B(\ti\nu_2\to \mu^-\ti\chi^+_1)$ in the $\ti\mu-\ti\tau$ mixing scenario.
%We have also found that the dependences of the decay
%branching ratios on the LFV parameters are similar as in the 
%$\tilde\mu-\tilde\tau$ mixing scenario. For instance, $B(\ti\nu_3\to
%\tau^-\ti\chi^+_1)$ shows a similar dependence on $R_{L13}$
%(where $R_{L13} \equiv M^2_{L,13}/((M^2_{L,11}+M^2_{L,22}+M^2_{L,33})/3)$) as 
%$B(\ti\nu_2\to \tau^-\ti\chi^+_1)$ on $R_{L23}$ (see Fig.2).
We have also found that the dependences of the decay
branching ratios on the LFV parameters are similar (including the excluded 
regions) to those in the $\tilde\mu-\tilde\tau$ mixing scenario. For instance, 
$B(\ti\nu_3\to\tau^-\ti\chi^+_1)$ shows a similar dependence on $R_{L13}$
($\equiv M^2_{L,13}/((M^2_{L,11}+M^2_{L,22}+M^2_{L,33})/3)$) to that of 
$B(\ti\nu_2\to \tau^-\ti\chi^+_1)$ on $R_{L23}$ shown in Fig.2.
%This is due to the fact that the experimental 
%limits on $B(\tau^- \to e^- \gamma)$ and $B(\tau^- \to \mu^- \gamma)$ are
%comparable and that the theoretical limits of the
%conditions (i) on the LFV parameters $A_{13}$ and $A_{31}$ are also
%similar to those on $A_{23}$ and $A_{32}$.
This is due to the fact that $Y_{E,11} \sim Y_{E,22} (\sim 0)$, that the experimental 
limits on $B(\tau^- \to e^- \gamma)$ and $B(\tau^- \to \mu^- \gamma)$ are
%comparable, and that the theoretical limits of the
%condition (i) on the LFV parameters $A_{13}$ and $A_{31}$ are also
%{\bf
comparable, and that the theoretical constraints on the LFV parameters
$A_{13}$ and $A_{31}$ from condition (i) are also
similar to those on $A_{23}$ and $A_{32}$.
%}

%-------------------------------------------------------------------
\subsection{$\tilde\nu$ production cross sections \label{sigma}}
%-------------------------------------------------------------------

It is to be noted that the production cross sections
%$\sigma(e^+e^-\to\sn_i\bar{\sn}_j)$ for $i\neq j$, $i,j=1,2,3$, are negligible 
%in the $\tilde \mu-\tilde\tau$ mixing scenario given in Table \ref{tab1}.
%This is due to the fact that in this scenario the t-channel chargino exchanges
%contribute significantly only to $\sigma_{33}$ (with
%$\tilde\nu_3\approx \tilde\nu_e$). 
$\sigma(e^+e^-\to\sn_i\bar{\sn}_j) \equiv \sigma_{ij}$ for $i\neq j$ with $i,j=1,2,3$ 
are negligible in the $\tilde \mu-\tilde\tau$ mixing scenario of Table \ref{tab1}.
This is due to the fact that in this scenario the t-channel chargino exchanges
contribute significantly only to $\sigma_{33}$ (with $\tilde\nu_3\approx \tilde\nu_e$).
Furthermore, the s-channel $Z$ boson
exchange cannot contribute to the cross section for $i\neq j$ because
of the vanishing $Z\tilde\nu_i\bar{\tilde\nu}_j$ couplings, which
is due to the unitarity of the mixing matrix $R^{\ti\nu}$.
%For the same reason the dependence of the cross sections
%$\sigma(e^+e^-\to\sn_i\bar{\sn}_i)$, $i=1,2,3$, on the LFV parameter
%$M^2_{L,23}$ is very weak in this scenario.\\
For the same reason the dependence of the cross sections
$\sigma_{ii}$ with $i=1,2,3$ on the LFV parameter
$M^2_{L,23}$ is very weak in this scenario.

In the $\tilde e-\tilde\tau$ mixing scenario, however, the 
influence of slepton generation mixing on
$\sigma(e^+e^-\to\sn_i\bar{\sn}_j)$ is completely 
different from that in the $\tilde\mu-\tilde\tau$ mixing scenario.
The reason is that in the $\ti e-\ti\tau$ mixing scenario the t-channel 
chargino exchanges 
%contribute significantly to the cross section
%$\sigma(e^+e^-\to\sn_i\bar{\sn}_j)$ for $i,j=1,2,3$,  
%yielding pronounced dependences of the cross sections on the
%$\tilde\nu_e-\tilde\nu_\tau$ mixing parameter $R_{L13}$.\\
contribute significantly to the cross sections $\sigma_{ij}$ for $i,j=$1,3, 
enhancing the cross sections strongly and 
yielding pronounced dependences of the cross sections on the
$\tilde\nu_e-\tilde\nu_\tau$ mixing parameter $M^2_{L,13}$.
Note that $\sn_1$ ($\sn_3$) is mainly $\sn_\tau$ ($\sn_e$) with a small 
$\sn_e$ ($\sn_\tau$) component. \\
%
%In Fig.6 we show the $R_{L13}$ dependence of 
%$\sigma(e^+e^-\to\sn_1\bar{\sn}_1)$, $\sigma(e^+e^-\to\sn_1\bar{\sn}_3)$
%($=\sigma(e^+e^-\to\sn_3\bar{\sn}_1)$) and
%$\sigma(e^+e^-\to\sn_3\bar{\sn}_3)$ at $\sqrt{s}=1$~TeV for
In Fig.6 we show the $R_{L13}$ dependence of $\sigma_{11}$, $\sigma_{13}$ 
($=\sigma_{31}$) and $\sigma_{33}$ at $\sqrt{s}=1$~TeV for 
longitudinal polarization of -90\% and 60\% of the electron and 
positron beam, respectively. All parameters other than $M^2_{L,13}$
are fixed as in our $\tilde e-\tilde\tau$ mixing scenario described 
in subsection \ref{BRs}.
%We remark that the production of a pair of sneutrinos with different masses 
%is a unique signal for LFV. 
%Its cross section can be as large as about 17fb in this scenario. 
We remark that the production of a pair of sneutrinos with different masses 
is a unique signal of LFV. 
This cross section $\sigma_{13}$ ($=\sigma_{31}$) can be as large as about 17fb 
in this scenario as can be seen in Fig.6. 
%{\bf
The analysis of \cite{Freitas:2004} shows that the sneutrino mass can be 
measured with an error of about 0.6 \% at ILC. In our case this corresponds to 
a mass uncertainty of about 2.5 GeV and hence the expected mass difference 
$m_{\sn_3} - m_{\sn_1}$ of about 10 GeV should be measurable at ILC.
%}

%------------------------------------------------------------------
\subsection{LFV contributions to collider signatures}
%------------------------------------------------------------------

Now we study the LFV contributions to signatures of sneutrino
production and decay at the ILC.
%
%In $e^+e^-$ collisions the sensitivity
%to probe $\tilde e-\tilde\tau$ generation mixing is in general larger
%compared to $\tilde\mu-\tilde\tau$ generation mixing, as  
%the sneutrino production cross sections are larger
%because of the sizable t-channel chargino contributions \cite{Nomura:2000zb}.
%
At $e^+e^-$ colliders the experimental sensitivity
to $\tilde e-\tilde\tau$ mixing is in general larger
than that to $\tilde\mu-\tilde\tau$ mixing because of the larger sneutrino 
production cross sections due to the sizable t-channel chargino exchange 
contributions \cite{Nomura:2000zb}.
Therefore, we first discuss signatures in the $\tilde e-\tilde\tau$
mixing scenario described in subsection \ref{BRs}.
We calculate the LFV contributions in sneutrino production and decay
to the rate of the signal events
\begin{equation}
e^\pm \tau^\mp+4{\rm jets}+/\!\!\!\!E\qquad {\rm and}\qquad
e^\pm \tau^\pm \tau^\mp+2{\rm jets}+/\!\!\!\!E~,
\label{eq:signal1}
\end{equation}
where $/\!\!\!\!E$ is the missing energy
\footnote{
Note that there are also contributions from LFV transitions in 
the production and decay of charged sleptons to the 
signals in (\ref{eq:signal1}).
}. 
We assume $\sqrt{s}=1$~TeV and a longidutinal polarization of -90\% and 60\% for 
the electron and positron beam, respectively.
From Table \ref{tab4} we see that
the dominant LFV contributions to the rate of the signal 
event $e^\pm \tau^\mp+4{\rm jets}+/\!\!\!\!E$ is obtained by 
\footnote{
We remark that the effect of sneutrino flavour oscillation
is negligible since in the considered scenario we have
$m^2_{\tilde\nu_i}-m^2_{\tilde\nu_j}\gg 
1/2(m_{\tilde\nu_i} \Gamma_{\tilde\nu_i}+
m_{\tilde\nu_j} \Gamma_{\tilde\nu_j})$ for $i>j$ \cite{Arkani-Hamed:1996au}.
Hence, the rate of the combined process factorizes into the product
of production cross section and decay branching ratios.
} 
\begin{eqnarray}
\sigma^{LFV}_1 &=&
\sigma_1 \cdot \sum_q B(\tilde\chi_1^+ \to q{\bar q}' \tilde\chi^0_1) \cdot 
\sum_q B(\tilde\chi_1^- \to q{\bar q}' \tilde\chi^0_1)~,\nonumber\\[2mm]
\sigma_1&=&2~\sum^3_{i=1}\sum^3_{j=1}~
\sigma(e^+e^- \to \tilde\nu_i \bar{\tilde\nu}_j)~
B(\tilde\nu_i \to \tau^- \tilde\chi^+_1)~
B(\bar{\tilde\nu}_j \to e^+ \tilde\chi^-_1)~,
\label{eq:sigma1}
\end{eqnarray}
which we calculate using the program SPheno \cite{Porod:2003um}.
%
%We obtain the cross sections $\sigma_{11}=14$fb, $\sigma_{22}=5$fb,
%$\sigma_{33}=99$fb, $\sigma_{12}=0$, $\sigma_{13}=10$fb, $\sigma_{23}=0$,
%where $\sigma_{ij}\equiv \sigma(e^+e^-\to\tilde\nu_i\bar{\tilde\nu}_j)$.
%The factor 2 in Eq.~(\ref{eq:sigma1}) is due
%to a summation over the charges of the involved leptons.
%Furthermore, we get $\sum_q B(\tilde\chi_1^\pm \to q{\bar q}'
%\tilde\chi^0_1)=0.67$ in this scenario.
%
The factor 2 in Eq.~(\ref{eq:sigma1}) is due
to the summation over the charges of the involved leptons.
For the cross sections $\sigma_{ij} \equiv 
\sigma(e^+e^-\to\tilde\nu_i\bar{\tilde\nu}_j)$ we obtain $\sigma_{11}=14$fb, 
$\sigma_{22}=5$fb, $\sigma_{33}=99$fb, $\sigma_{13}=\sigma_{31}=10$fb, 
$\sigma_{12}=\sigma_{21}=\sigma_{23}=\sigma_{32}=0$.
Further we get $\sum_q B(\tilde\chi_1^\pm \to q{\bar q}'
\tilde\chi^0_1)=0.67$ in this scenario.
As a result, the corresponding cross section  
that gives rise to $e^\pm \tau^\mp+4{\rm jets}+/\!\!\!\!E$ is $\sigma^{LFV}_1=6.6$fb.
Note however that lepton flavour conserving (LFC) transitions also can 
contribute to the final state $e^\pm \tau^\mp+4{\rm jets}+/\!\!\!\!E$.
The dominant contribution of this type stems from 
the reaction $e^+e^-\to \tilde\nu_3 \bar{\tilde\nu}_3$, followed
by the LFC decays $\tilde\nu_3 \to e^- \tilde\chi^+_1$ and 
$\tilde\chi^+_1 \to \tau^+ \nu \tilde\chi^0_1$, with
the branching ratio
$B(\tilde\chi^+_1 \to \tau^+ \nu \tilde\chi^0_1)=0.1$.
The other $\bar{\tilde\nu}_3$ decays as
$\bar{\tilde\nu}_3 \to \bar\nu\tilde\chi^0_3$, where subsequently
the $\tilde\chi^0_3$ decays into $Z \tilde\chi^0_2$ (with a branching ratio 
(BR) of 6\%) and into $W^\mp \tilde\chi^\pm_1$ (with BR of 9.3\%). 
Both decay chains lead to a final state
containing 4jets plus missing energy, where
for the various decay branching ratios we have $\sum_qB(Z\to q{\bar q})=0.7$, 
$\sum_q B(\tilde\chi^0_2\to q{\bar q}\tilde\chi^0_1)=0.62$ and
$\sum_q B(W\to q{\bar q}')=0.68$.
Taking into account the branching ratios in the decay chains and 
summing over the charges of the involved particles, we find that
the corresponding cross section is $\sigma^{LFC}_1=0.033$fb, 
%which is more than two orders of magnitude smaller than $\sigma^{LFV}_1$. 
which is two orders of magnitude smaller than $\sigma^{LFV}_1$. 
One can expect that the kinematical distributions of LFV and LFC
contributions are quite different. Hence suitable kinematical 
cuts would further enhance the ratio of LFV to LFC contributions.\\  
Now we consider the LFV transitions which give rise to the
event $e^\pm \tau^\pm \tau^\mp+2{\rm jets}+/\!\!\!\!E$.
The dominant LFV contributions to the rate of the events 
is given by 
\begin{eqnarray}
\sigma^{LFV}_2 &=&\sigma_2 \cdot B(H^+ \to \tau^+ \nu) \cdot 
\sum_q B(\tilde\chi_1^- \to q{\bar q}' \tilde\chi^0_1)~,\nonumber\\[2mm]
\sigma_2 &=&2~\sum^3_{i=1}\sum^3_{j=1}\sum^3_{k=1}~
\sigma(e^+e^-\to \tilde\nu_i \bar{\tilde\nu}_j)~
B(\tilde\nu_i\to\tilde\ell_k^- H^+)\nonumber\\[3mm]
&&{}\times\left[
B(\bar{\tilde\nu}_j \to e^+ \tilde\chi^-_1)~
B(\tilde\ell^-_k \to  \tau^- \tilde\chi^0_1)+
B(\bar{\tilde\nu}_j \to \tau^+\tilde\chi^-_1)~
B(\tilde\ell^-_k \to e^-\tilde\chi^0_1)
\right],
\label{eq:sigma2}
\end{eqnarray}
where $B(H^+ \to \tau^+\nu)  \approx 1$, $B(\ti\ell^-_{1,2,3}\to
e^-\ti\chi^0_1)=(0,0,0.89)$ and $B(\ti\ell^-_{1,2,3}\to
\tau^-\ti\chi^0_1)=(0.38,0,0.006)$.
We obtain $\sigma^{LFV}_2=6.7$fb.
%
%Also in this case there are contributions to
%$e^\pm \tau^\pm \tau^\mp+2{\rm jets}+/\!\!\!\!E$
%from reactions that would contribute in the LFC case as well.
%
Also in this case there are LFC contributions to
$e^\pm \tau^\pm \tau^\mp+2{\rm jets}+/\!\!\!\!E$.
The dominant one is due to the reaction 
$e^+e^-\to \tilde\nu_1 \bar{\tilde\nu}_1$, 
%followed by the decays $\tilde\nu_1\to \tau^- \tilde\chi^+_1$ 
followed by the LFC decays $\tilde\nu_1\to \tau^- \tilde\chi^+_1$ 
and $\bar{\tilde\nu}_1\to \tau^+ \tilde\chi^-_1$, where subsequently
%
%one chargino decays hadronically and the other chargino decays as
%$\tilde\chi^+_1\to e^+\nu\tilde\chi^0_1$, with 
%$B(\tilde\chi^+_1 \to e^+ \nu %\tilde\chi^0_1)=0.12$.
%
one chargino decays hadronically and the other chargino decays as
$\tilde\chi^{\pm}_1 \to e^{\pm} \nu \tilde\chi^0_1$ with 
$B(\tilde\chi^{\pm}_1 \to e^{\pm} \nu \tilde\chi^0_1)=0.12$.
The corresponding cross section for 
this reaction is $\sigma^{LFC}_2=0.31$fb, which is about a factor 1/20 smaller
than $\sigma^{LFV}_2$.

We have also studied the LFV contributions to the rate of
the signal events $\mu^\pm \tau^\mp+4{\rm jets}+/\!\!\!\!E$ and
$\mu^\pm \tau^\pm \tau^\mp+2{\rm jets}+/\!\!\!\!E$ in the $\ti\mu-\ti\tau$
%mixing scenario (see Table \ref{tab1}).
%Taking the branching ratios, using Eqs.~(\ref{eq:sigma1}) and
%(\ref{eq:sigma2}) with $e$ replaced by $\mu$ we obtain
%the corresponding LFV cross sections $\sigma^{LFV}_1=0.14$fb and $\sigma^{LFV}_2=0.23$fb.
%Similarly, for the LFC cross sections we obtain $\sigma^{LFC}_1=0.0034$fb 
%and $\sigma^{LFC}_2=0.13$fb.
%
mixing scenario of Table \ref{tab1}.
Taking the branching ratios of Table \ref{tab3}, and using Eqs.~(\ref{eq:sigma1}) and
(\ref{eq:sigma2}) with $e$ replaced by $\mu$ in the branching ratios, we obtain
the corresponding LFV cross sections $\sigma^{LFV}_1=0.14$fb and 
$\sigma^{LFV}_2=0.23$fb, respectively.
Similarly, for the corresponding LFC cross sections we obtain $\sigma^{LFC}_1=0.0034$fb 
and $\sigma^{LFC}_2=0.13$fb, respectively.

%The examples above show that the LFV terms can contribute significantly to 
%event rates at linear colliders.
The examples above show that the LFV processes (including the LFV bosonic 
decays also) can contribute significantly to signal 
event rates at linear colliders.
%This suggests that one should include the LFV parameters in the determination
%of the basic parameters at colliders.
This strongly suggests that one should take into account the 
possibility of the significant contributions of both the LFV 
fermionic and bosonic decays in the sneutrino search and 
should also include the LFV parameters in 
the determination of the basic SUSY parameters at colliders.
It is clear that detailed Monte Carlo studies taking into account
background and detector simulations are necessary
\footnote{It has been shown \cite{Hisano:1998wn} that the background stemming
from the decay of $\tau$ into $e/\mu$, can be reasonably suppressed by 
suitable cuts on the $e/\mu$ energy 
and/or on the $e/\mu$ impact parameter.}.
However, this is beyond the scope of the present article.

%{\bf
Before closing this section we remark that 
the slepton generation mixings as discussed here could also significantly 
affect the production and decays of charged sleptons and that 
complex phases of the LFV parameters, such as $M^2_{L,23}$, might have 
an important influence on the sneutrino phenomenology \cite{Bartl:2002bh}. 
%These aspects will be discussed in forthcoming papers.
These topics deserve detailed studies and will be treated in 
forthcoming papers.
%}

%-------------------------------------------------------------------
\section{Summary\label{summary}}
%-------------------------------------------------------------------

We have studied sneutrino decays in the cases of $\tilde\mu-\tilde\tau$ 
and $\ti e-\tilde\tau$ mixings in the slepton sector. 
In the case of $\tilde e-\tilde\tau$ mixing we have obtained similar results to 
those in the $\tilde\mu-\tilde\tau$ mixing case since the experimental 
limit on $B(\tau^- \to e^- \gamma)$ is of the same order of magnitude as the 
one on $B(\tau^- \to \mu^- \gamma)$. The theoretical limits from the vacuum
stability conditions on the LFV parameters $A_{13}$ and $A_{31}$ are also 
similar to those on $A_{23}$ and $A_{32}$. 
In the $\tilde e-\tilde\mu$ mixing case the LFV effect is much smaller
because the experimental limit on $B(\mu^- \to e^- \gamma)$ is much stronger than 
those on $B(\tau^- \to \mu^- \gamma)$ and $B(\tau^- \to e^- \gamma)$ and the 
theoretical limits from the vacuum stability conditions on the LFV parameters $A_{12}$ and 
$A_{21}$ are also much stronger than those on $A_{23}, A_{32}, A_{13}$ and $A_{31}$. \\

To conclude, we have performed a systematic study of sneutrino production and 
decays including both fermionic and bosonic decays in the general MSSM with 
slepton generation mixings.
%We have shown that LFV sneutrino decay branching ratios 
%can be quite large due to LFV in slepton sector 
We have shown that LFV sneutrino production cross sections and 
LFV sneutrino decay branching ratios
can be quite large due to LFV in the slepton sector 
(i.e. due to slepton generation mixing) in a significant part of the MSSM parameter 
space despite the very strong experimental limits on LFV processes.
This could have an important impact on the search for sneutrinos and the 
MSSM parameter determination at future colliders, such as LHC, ILC, 
CLIC and muon collider.

%------------------------------------------------------------------------
\section*{Acknowledgements}
%------------------------------------------------------------------------
K.H. thanks H. Haber for useful comments and continuous encouragement.
W.P. thanks E. Arganda and M. J. Herrero
for providing the code for the calculation of the branching ratios 
$B(\tau\to 3\ell)$.
This work is supported by the 'Fonds zur F\"orderung der
wissenschaftlichen Forschung' (FWF) of Austria, project No. P18959-N16
and by the German Ministry of Education and Research (BMBF) under
contract 05HT6WWA.
The authors acknowledge support from EU under the MRTN-CT-2006-035505
network programme.

%------------------------------------------------------------------------
% references
%------------------------------------------------------------------------

\newpage

%------------------------------------------------------------------------
% Figures Captions
%------------------------------------------------------------------------
\begin{flushleft}
{\Large \bf Figure Captions} \\
\end{flushleft}

\noi
{\bf Figure 1}: 
%
%Contours of (a) $B(\sn_2 \to \tau^-  \chp_1)$ 
Contours of the LFV decay branching ratios (a) $B(\sn_2 \to \tau^-  \chp_1)$ 
and (b) $B(\sn_2 \to \sl^-_1  H^+)$ in the $\mu-M_2$ 
plane for our $\ti\mu-\ti\tau$ mixing scenario. All parameters other than 
$\mu$ and $M_2$ are fixed as in Table \ref{tab1}. 
The region with no solid contour-lines is excluded by the conditions (i) to 
(v) given in the text; negative $\mu$ region is excluded by the limit on 
$\Delta a_\mu^{SUSY}$ in (iii). The dashed and dash-dotted lines in (a) 
show contours of $B(\tau^- \to \mu^-  \gamma)$ and $\Delta a_\mu^{SUSY}$, 
respectively. Note that (iii) requires $1.09 \times 10^{-9} < 
\Delta a_\mu^{SUSY} < 4.65 \times 10^{-9}$.

\noi 
{\bf Figure 2}:
$R_{L23}$ dependence of $\sn_2$ decay branching ratios for our $\ti\mu-\ti\tau$ mixing scenario, where 
all basic parameters other than $M^2_{L,23}$ are fixed 
as in Table \ref{tab1}. The shown range of 
$R_{L23}$ is the whole range allowed by the conditions 
(i) to (v) given in the text.

\noi 
{\bf Figure 3}: 
%Contours of (a) $B(\sn_2 \to \tau^-  \chp_1)$ 
Contours of the LFV decay branching ratios (a) $B(\sn_2 \to \tau^-  \chp_1)$ 
and (b) $B(\sn_2 \to \sl^-_1  H^+)$ 
in the $R_{L23}-R_{A23}$ plane for our $\ti\mu-\ti\tau$ mixing scenario, where
all basic parameters other than $M^2_{L,23}$ and 
$A_{23}$ are fixed as in Table \ref{tab1}. 
The region with no solid contours is excluded by the conditions (i) to (v) 
given in the text. The dashed lines in (a) show contours of $B(\tau^- \to \mu^- \gamma)$.

\noi 
{\bf Figure 4}: 
Contours of the LFV decay branching ratio $B(\sn_1 \to \sl^-_2  H^+)$ 
in the $R_{E23}-R_{A32}$ plane for our $\ti\mu-\ti\tau$ mixing scenario, 
where all basic parameters 
other than $M^2_{E,23}$ and $A_{32}$ are fixed as in Table \ref{tab1}. 
The region with no solid contours is excluded by the conditions 
(i) to (v) given in the text. The dashed lines show contours of 
$B(\tau^- \to \mu^- \gamma)$. 

\noi 
{\bf Figure 5}: 
Scatter plot of the LFV decay branching ratios 
$B(\sn_2 \to \sl^-_1  H^+)$ versus $B(\tau^- \to \mu^- \gamma)$ for our 
$\ti\mu-\ti\tau$ mixing scenario with the parameters 
$M_2, \mu, R_{L23}, R_{E23}, R_{A23}$ and $R_{A32}$ varied in the ranges 
$0<M_2<1000$~GeV, $|\mu|<1000$~GeV, $|R_{L23}|<0.1, |R_{E23}|<0.2, |R_{A23}|<2.5$ and 
%$|R_{A32}|<2.5$, respectively, satisfying the conditions (i) to (v) given in the text. 
$|R_{A32}|<2.5$, satisfying the conditions (i) to (v) given in the text. 
All parameters other than $ M_2, \mu, M^2_{L,23}, M^2_{E,23}, 
A_{23} $ and $A_{32}$ are fixed as in Table \ref{tab1}.

\noi 
{\bf Figure 6}: 
%Production cross sections 
%$\sigma(e^+e^-\to\sn_1\bar{\sn}_1)$ (dotted line),
%$\sigma(e^+e^-\to\sn_1\bar{\sn}_3)$ ($=\sigma(e^+e^-\to\sn_3\bar{\sn}_1)$) (solid line) and
%$\sigma(e^+e^-\to\sn_3\bar{\sn}_3)$ (dashed line)
%as a function of $R_{L13}$ for our $\ti e-\ti\tau$ mixing scenario at
%$\sqrt{s}=1$~TeV. 
Production cross sections 
$\sigma(e^+e^-\to\sn_1\bar{\sn}_1)$ (dotted line),
$\sigma(e^+e^-\to\sn_1\bar{\sn}_3)$ ($=\sigma(e^+e^-\to\sn_3\bar{\sn}_1)$) (solid line) and
$\sigma(e^+e^-\to\sn_3\bar{\sn}_3)$ (dashed line) at $\sqrt{s}=1$~TeV 
as a function of $R_{L13}$ for our $\ti e-\ti\tau$ mixing scenario. 
All parameters other than $M^2_{L,13}$ are fixed 
%as in $\ti e-\ti\tau$ mixing scenario
as in the $\ti e-\ti\tau$ mixing scenario described in subsection \ref{BRs}. 
The $R_{L13}$ region shown is allowed by the conditions (i) to (v) given in
the text.

\newpage
%
%------------------------------------------------------------------------
% Figures
%------------------------------------------------------------------------
%
% Figure 1 --------------------------------------------------------------
%
\begin{figure}[!htb] 
\begin{center}
%\scalebox{0.6}[0.76]{\includegraphics{fig1a.eps}} \\
%\vspace{2mm}
%
%\scalebox{0.66}[0.76]{\includegraphics{fig1a.eps}} \\
%\vspace{2mm}
%\scalebox{0.6}[0.76]{\includegraphics{fig1b.eps}} \\
%
\scalebox{0.7}[0.8]{\includegraphics{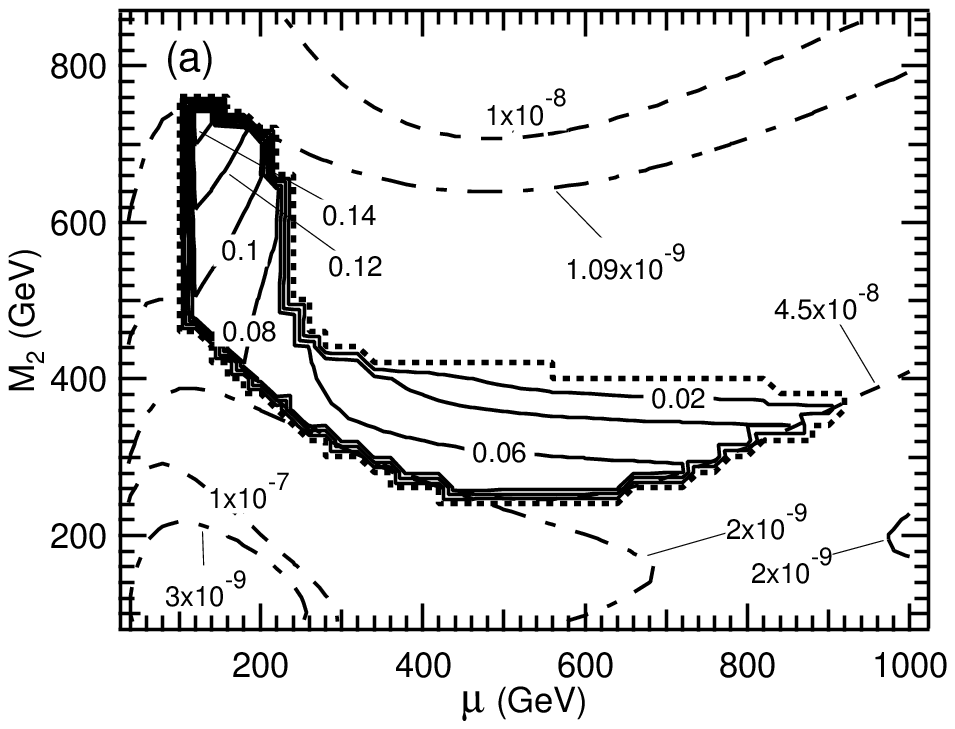}} 
%\vspace{2mm}
\scalebox{0.64}[0.81]{\includegraphics{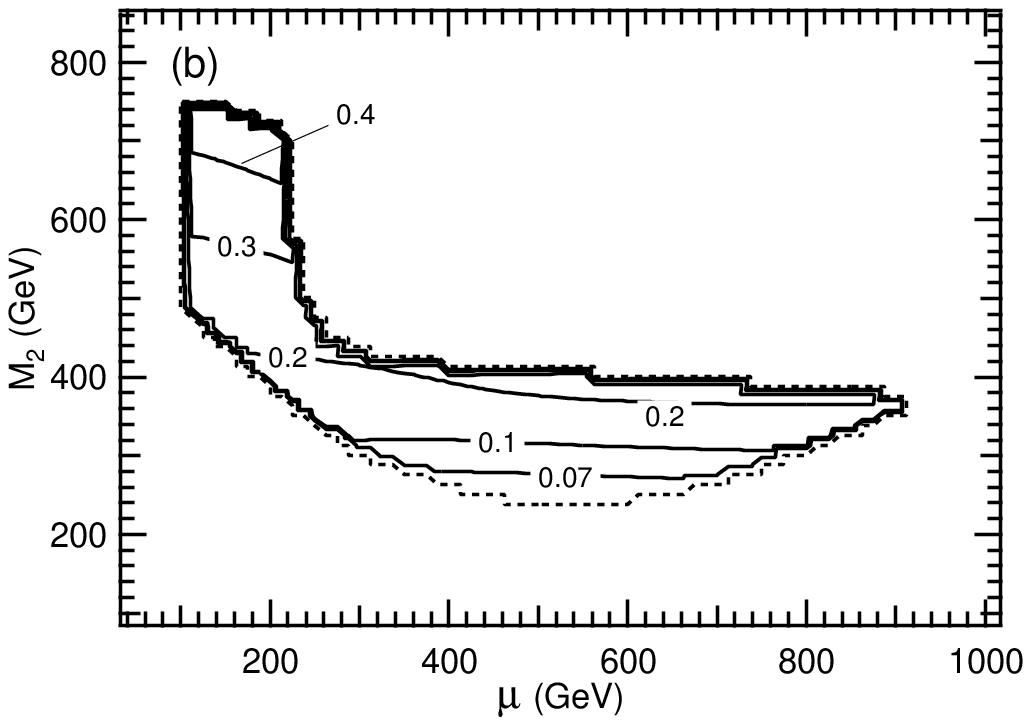}} \\
%
%\vspace{5mm}
{\LARGE \bf Fig.1}
\end{center}
\end{figure}

%\newpage
%
% Figure 2 ------------------------------------------------------------
%
\begin{figure}[!htb] 
\begin{center}
%\scalebox{0.8}[1.0]{\includegraphics{fig2.eps}} \\ 
\scalebox{0.7}[0.9]{\includegraphics{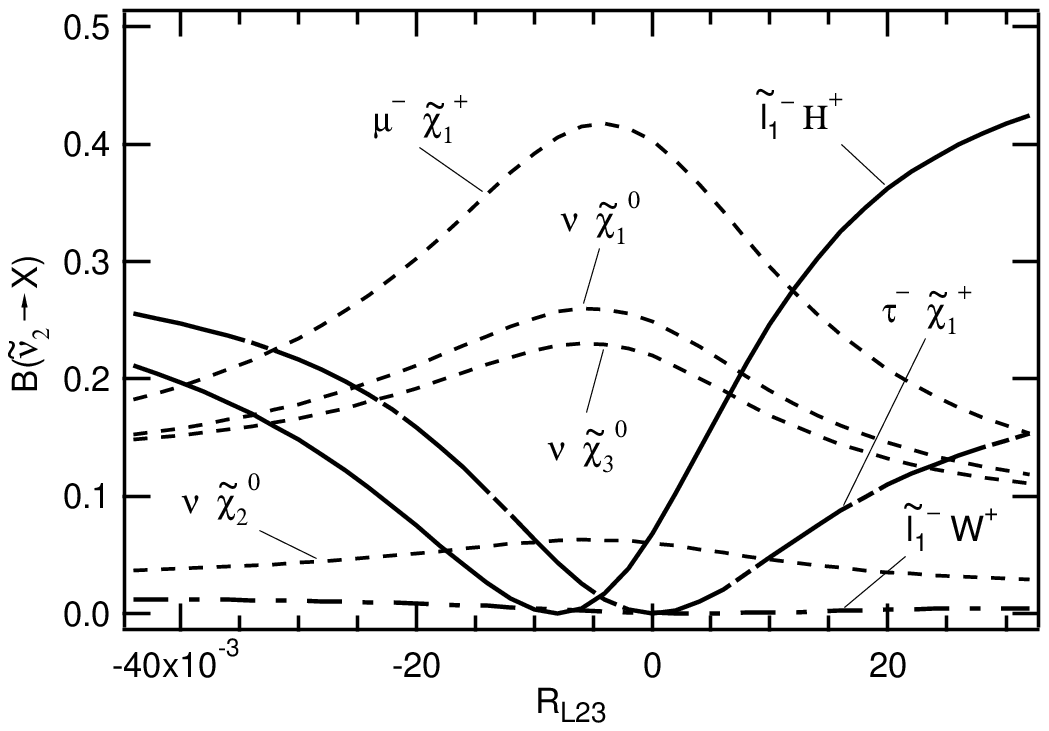}} \\ 
%\vspace{-5mm}
%\scalebox{0.95}[1.1]{\includegraphics{fig3b.eps}} 
\end{center}
\end{figure}
\vspace{-10mm}
\begin{center}
{\LARGE \bf Fig.2}
\end{center}

\newpage
%
% Figure 3 ------------------------------------------------------------
%
\begin{figure}[!htb] 
\begin{center}
%\scalebox{0.55}[0.7]{\includegraphics{fig3a.eps}} 
%\vspace{3mm}
%\scalebox{0.55}[0.7]{\includegraphics{fig3b.eps}} 
\scalebox{0.65}[0.85]{\includegraphics{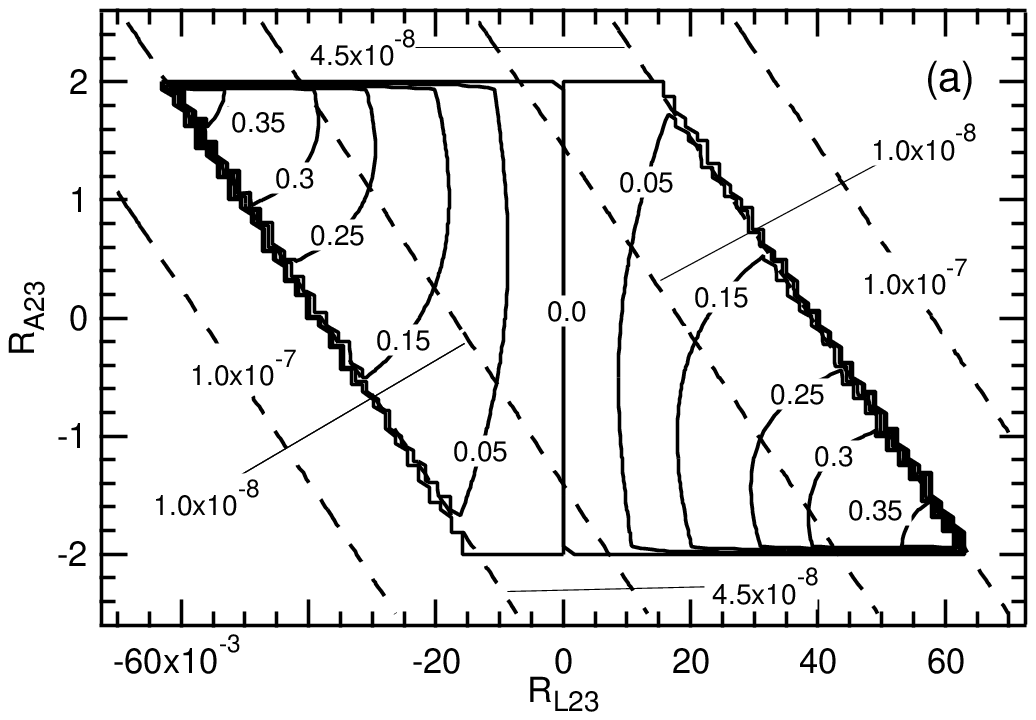}} 
%\vspace{3mm}
\scalebox{0.65}[0.85]{\includegraphics{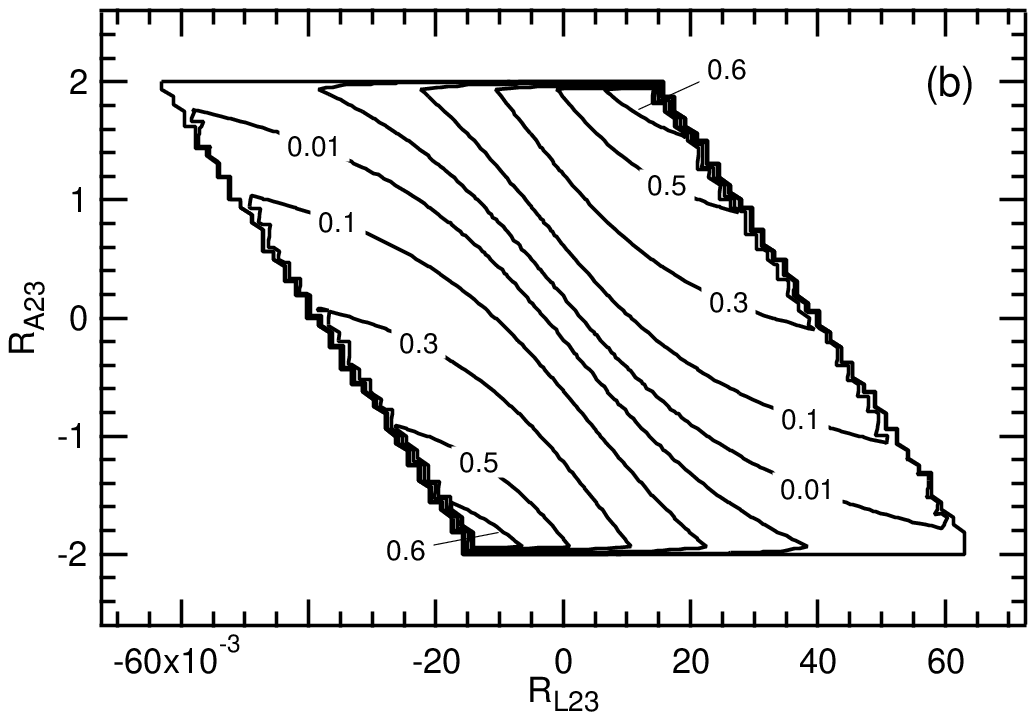}} 
\vspace{-10mm}
\end{center}
\end{figure}

\begin{center}
{\LARGE \bf Fig.3}
\end{center}

%\newpage
%
% Figure 4 ------------------------------------------------------------
%
\begin{figure}[!htb] 
\begin{center}
\scalebox{0.7}[1.0]{\includegraphics{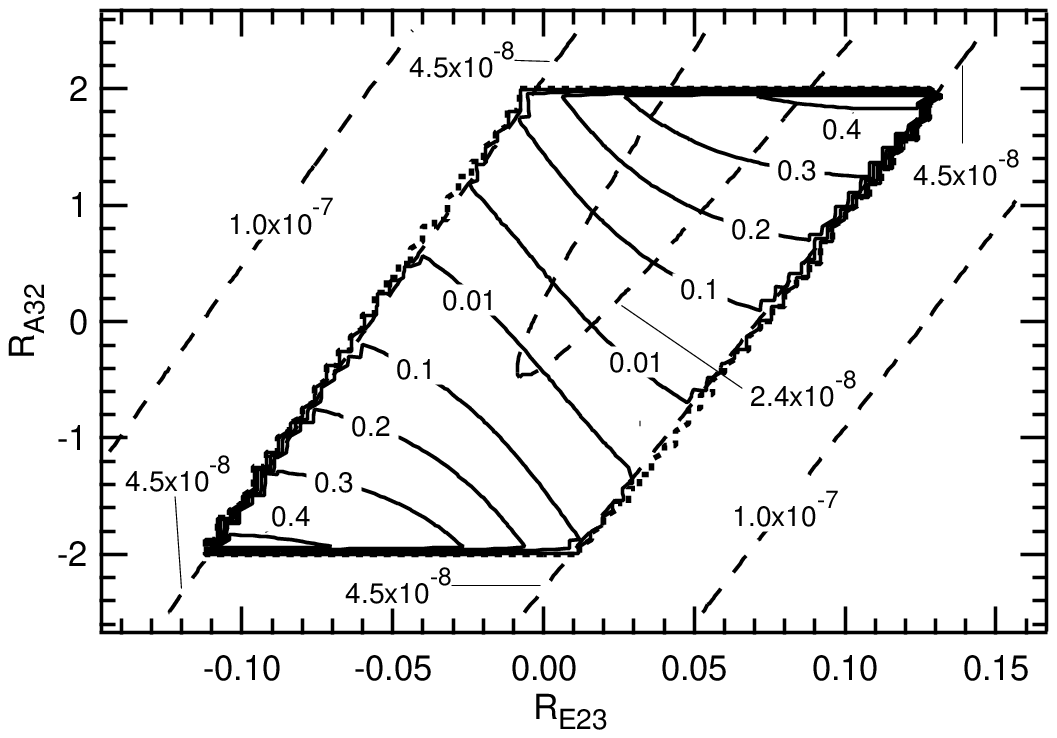}} \\
%\vspace{5mm}
{\LARGE \bf Fig.4}
\end{center}
\end{figure}
%

%\newpage
%
% Figure 5 ------------------------------------------------------------
%
\begin{figure}[!htb] 
\begin{center}
\scalebox{0.85}[1.05]{\includegraphics{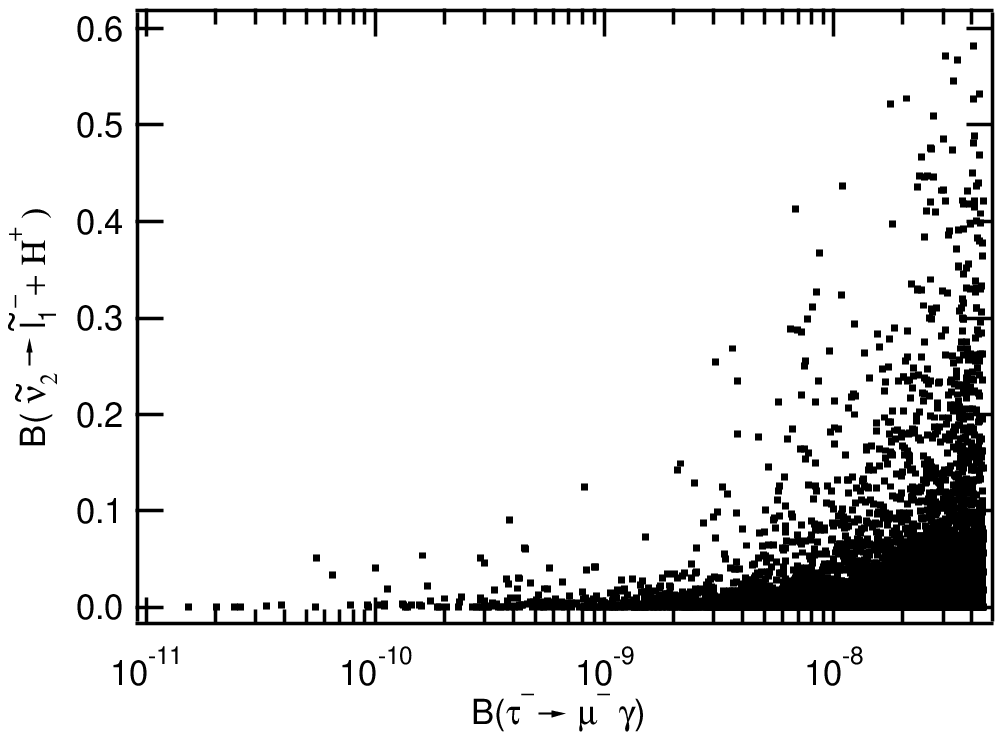}} \\ 
%\vspace{5mm}
{\LARGE \bf Fig.5}
\end{center}
\end{figure}
%
%\newpage
%
% Figure 6 ------------------------------------------------------------
%
\begin{figure}[!htb] 
\begin{center}
%\scalebox{1.7}[1.5]{\includegraphics{fig6.eps}} \\ 
\hspace{25mm}
\scalebox{1.3}[1.1]{\includegraphics{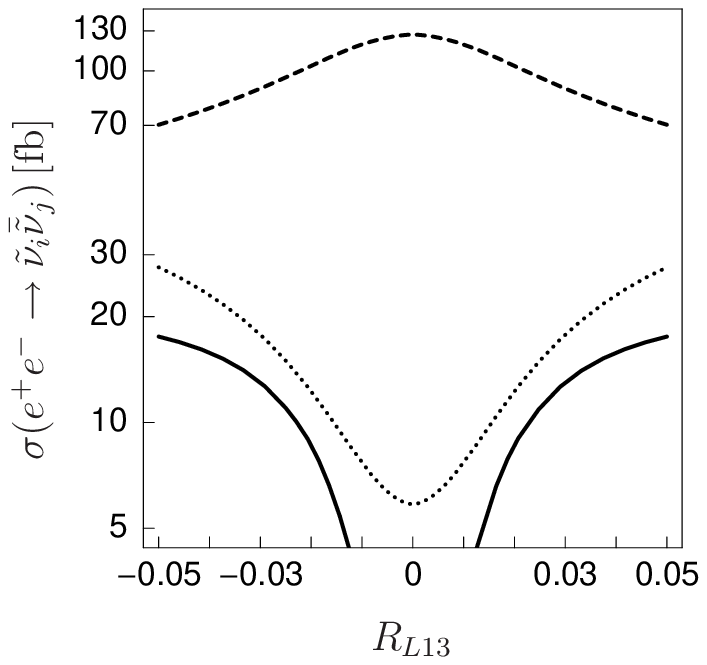}} \\ 
\vspace{5mm}
{\LARGE \bf Fig.6}
\end{center}
\end{figure}
%

%----------------------------------------------------------------------
\end{document}